\documentclass[nofootinbib,ps,reprint,prl,twocolumn]{revtex4-1}
\usepackage{amsmath}
\usepackage{amsfonts}
\usepackage{amssymb}
\usepackage{graphicx}
\usepackage{bbold}
\usepackage[colorlinks,linkcolor=blue,anchorcolor=blue,citecolor=blue,urlcolor=black]{hyperref}
\usepackage{mathrsfs}
\usepackage{dcolumn}
\usepackage{bm}
\usepackage{epsfig}
\usepackage[version=3]{mhchem}%
\newcommand{\xpt}{\mathbf{x}}
\newcommand{\ypt}{\mathbf{y}}
\newcommand{\xipt}{\mathbf{\xi}}
\newcommand{\kvec}{\mathbf{k}}
\newcommand{\Lmat}{\mathbf{L}}
\def\be{\begin{equation}}
\def\ee{\end{equation}}
\begin{document}
\title{Scarring in classical chaotic dynamics with noise }
\author{%Kensuke Yoshida$^{1,2}$, Hajime Yoshino$^1$, Akira Shudo$^1$,  and Domenico Lippolis$^2$
Domenico Lippolis$^1$, Akira Shudo$^2$, Kensuke Yoshida$^{2}$, and Hajime Yoshino$^2$
 }
%\email{domenico@ujs.edu.cn}
%\altaffiliation{URL: www.phy.pku.edu.cn/$\sim$yfxiao/}
\affiliation{
$^1$Institute for Applied Systems Analysis, Jiangsu University, Zhenjiang 212013, China
\\
$^2$Department of Physics, Tokyo Metropolitan University, Minami-Osawa, Hachioji 192-0397, Japan
} 
%\affiliation{

\date{\today}

\begin{abstract}
      
We report the numerical observation of scarring, that is enhancement of probability density around unstable periodic orbits of 
a chaotic system, in the eigenfunctions of the classical Perron-Frobenius operator of noisy Anosov (`cat') maps, as well
as in the noisy Bunimovich stadium. A parallel is drawn between classical and quantum scars, based on the unitarity or non-unitarity of the respective propagators. 
For uniformly hyperbolic systems such as the cat map, we provide a mechanistic explanation for the classical phase-space localization detected, based 
on the distribution of finite-time Lyapunov exponents, and the interplay of noise with deterministic dynamics.
Classical scarring can be measured by studying autocorrelation functions and their power spectra.

\end{abstract}

\pacs{42.55.Sa, 42.65.Sf, 05.45.-Gg, 05.45.-a}
\maketitle

\textit{Introduction.} In the realm of classical- and quantum chaos, phase-space densities tend to mix,
due to the stretching and folding action of the dynamics. 
As a result, every form of localization is an anomaly
to the expected `random' behavior of Hamiltonians,
propagators, wavefunctions, and various observables.

Examples in quantum mechanics include dynamical localization for a kicked rotor~\cite{cas79}, which 
can be related to the Anderson localization of a tight-binding model~\cite{GFP82},
opening-induced phase-space localization~\cite{Ketz18},
and probability density enhancement around unstable periodic orbits of the underlying
classical system. The latter is known as scarring~\cite{Hel84}, and it has drawn % elicited 
a fair amount of attention since its first discovery in the quantum Bunimovich stadium billiard.
Scars, the regions of enhanced probability density, have been ascribed to 
constructive interference around periodic orbits~\cite{Bogo88}. 
Dismissed for a while as transients of no effect on the long-term properties 
of a closed chaotic system subject to thermalization~\cite{Srednicki}, scars were brought back into
the spotlight by recent numerical
evidence of ergodicity breaking in many-body 
systems~\cite{MBScars,%PapicLong,NonintScars,
ConstrScars,PerfScars,ExactScars,
MagXYScars,
%MagScars,%RevivScars,
FractScars}. %Papic20}.
Further theoretical and experimental work has extended the notion of scarring to 
regular dynamics\cite{SuperBogo,LebScars,Dietz08}, to integrable systems
 with disorder~\cite{StrongScars16,ContScars17,Lissa19}, of interest in cold atoms and 
condensed matter, as well as to relativistic Dirac billiards~\cite{QRS,DirScars}.

In the present paper, 
we report scarring  
%that is probability density  enhancement  near unstable
%periodic orbits, 
%which we  call 
in the  eigenfunctions of the \textit{classical} Perron-Frobenius evolution operator~\cite{DasBuch}
with background noise,
for two paradigmatic models of  chaos.
The observations presented here suggest that
%support the more general idea that  
%stretches over the last decade,
%according to which 
quantum localization in chaos does not 
exclusively arise from interference, 
but is also a classical effect. 
%Here, in particular, 

The noiseless Perron-Frobenius operator
%defined as ~\cite{DasBuch}%\footnote{better to replace~(\ref{PF}) with the Fokker-Planck operator from the start?}
\begin{equation}
{\cal{L}}^t\rho(\xpt) = \int dx_0\,\delta\left(\xpt-f^t(\xpt_0)\right)\rho(\xpt_0)
\,.
\label{PF}
\end{equation}
transports an initial phase-space density of trajectories $\rho(\xpt)$ through the
flow $f^t(\xpt)$, that is the solution of the equations of motion,
to a new density. The Perron-Frobenius operator is linear, and 
its spectral properties depend in general on the space of
functions it acts upon~\cite{Dolgo98,BKL02,Liverani04,Mark04,Rugh92,CrawCar}. It is a formal solution to the 
Liouville equation $\partial_t\rho + \nabla\cdot (\rho \mathbf{v}) = 0$,
where $\dot{\xpt}=\mathbf{v}(\xpt)$ is the dynamical system in exam. %\footnote{write something on the space of function we select and the spectral properties e.g. discrete spectrum with isolated leading eigenvalue?}.    
In chaotic Hamiltonian systems with no escape, 
the Perron-Frobenius spectrum has an isolated, unitary eigenvalue, 
whose (`leading') eigenfunction is uniform in the phase space, and it is
called natural measure or invariant density~\cite{GaspBook}.
The natural measure is the weight to every phase space average, and, as such,
its successful determination 
%amounts to being able 
enables us to evaluate 
any long-term averaged observable under the ergodicity assumption,
thus solving the problem of statistical mechanics. %\footnote{perhaps ask Prof. Shudo to review the previous paragraph. Any heresies?}. 
Here, instead, we focus on the other, `subleading' eigenfunctions of the 
Perron-Frobenius spectrum, whose eigenvalues yield the decay rate 
of any initial density to %the steady state, that is 
the natural measure.
%To convey this idea, we 
In a suitable functional space, we can 
%Let us formally 
expand the evolution of a density as
\begin{equation}
{\cal{L}}^t\rho(\xpt) = \sum_n a_n\mathrm{e}^{-\gamma_nt}\phi_n(\xpt) + \sum_ma_m(t)\psi_m(\xpt)
\,
\label{Lexp}
\end{equation} 
where
$\phi_0(\xpt)=1$, $\gamma_0=0$,  %are the natural measure and the 
%log of its eigenvalue 
%escape rate respectively,
%while the %$\phi_1(\xpt)$ and $\gamma_1$  
while the summation over $n$ is an expansion over the 
%rest of the 
eigenfunctions, all decaying 
with rates $\gamma_n$ increasing with $n$, and the $\sum_m$ represents Jordan blocs, since the Perron-Frobenius 
operator is in general non-diagonalizable [its spectrum also has a continuous part, 
neglected in~(\ref{Lexp})]. In particular, $\phi_1(\mathbf{x})$ is hereafter referred to as second eigenfunction
of the evolution operator.

In reality, every physical system experiences noise in some form, which is modeled as a random variable $\xipt(t)$
in the equations of motion, $\dot{\xpt}=\mathbf{v}(\xpt) + \xipt(t)$ . If the noise is assumed as Gaussian-distributed and uncorrelated, 
the Liouville equation above acquires a diffusion term, say $D\nabla^2\rho$ 
%($\langle\xipt(t)\xipt(t+\tau)\rangle=2D\delta(\tau)$),
($D$ is the noise amplitude or variance of $\xipt$), 
and is known as Fokker-Planck equation. Its formal 
solution is a path (`Wiener') integral~\cite{Risk96}, whose kernel can be regarded as an evolution operator
analogous to the Perron-Frobenius in Eq.~(\ref{PF}), but with a finite-width distribution instead of 
the delta function. 
%In the discrete-time systems considered here, the Fokker-Planck operator 
%transforms densities as\footnote{This is never the operator we actually use. Either we use a noisy 
%Perron-Frobenius operator realized with a Monte-Carlo method, or the one~(\ref{FourOp}) in Fourier space} 
%\begin{equation}
%{\cal{L}}^t_{FP}\rho(x) = \int dx_0\,\mathrm{e}^{-\left(x-f^t(x_0)\right)^2/4D}\rho(x_0)
%\,.
%\label{FP}
%\end{equation}
%The spectral properties of ${\cal{L}}^t_{FP}$ and ${\cal{L}}^t$ are similar\footnote{really?}. 
%As the kernel of the Fokker-Planck operator in Eq.~(\ref{FP}) suggests,
Noise smears out densities, and thus it balances contractions  
from the deterministic, chaotic dynamics. Distributions of 
trajectories,  as well as eigenfunctions of the noisy Perron-Frobenius operator, are
then expected to be smooth. 
\begin{figure}[tbh!]
\centerline{
\includegraphics[width=4.8cm]{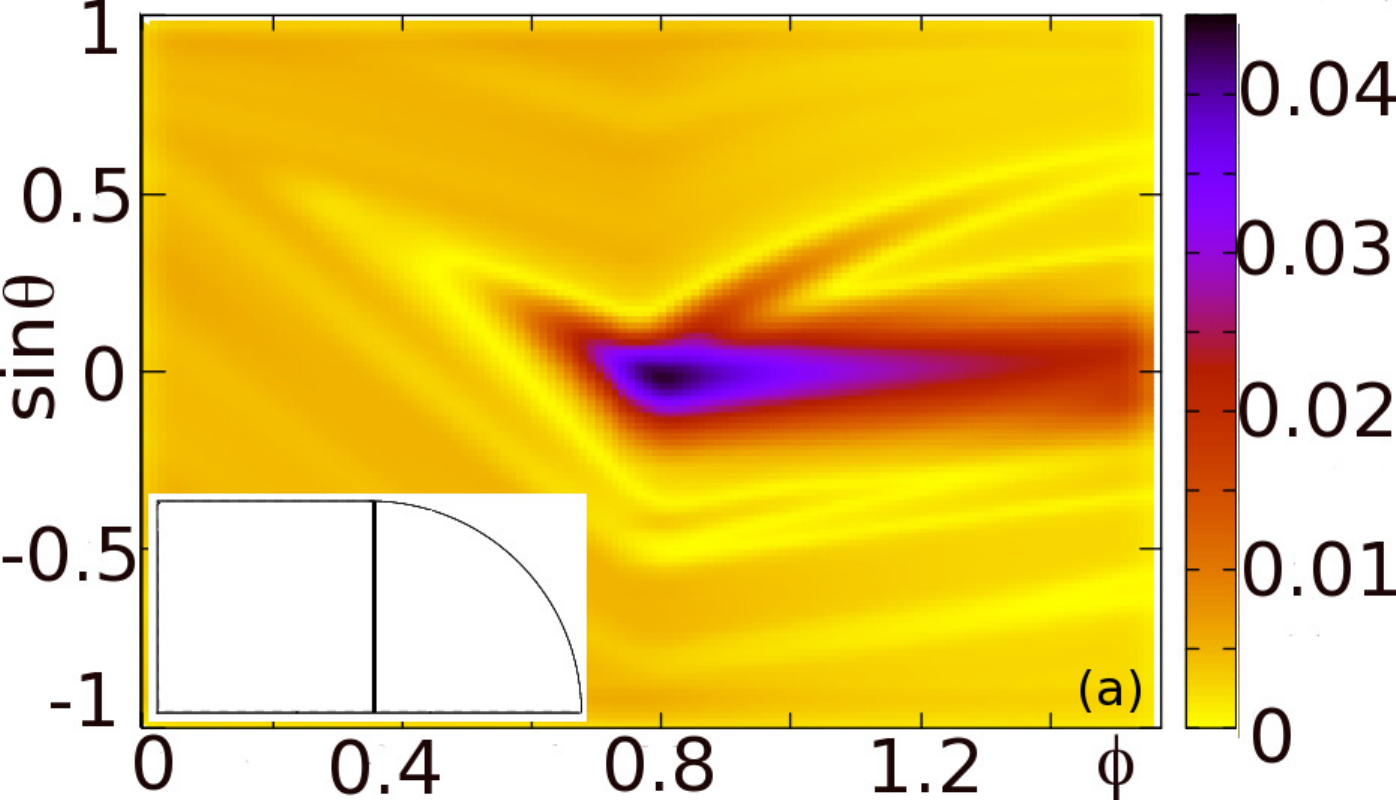}
\includegraphics[width=3.9cm]{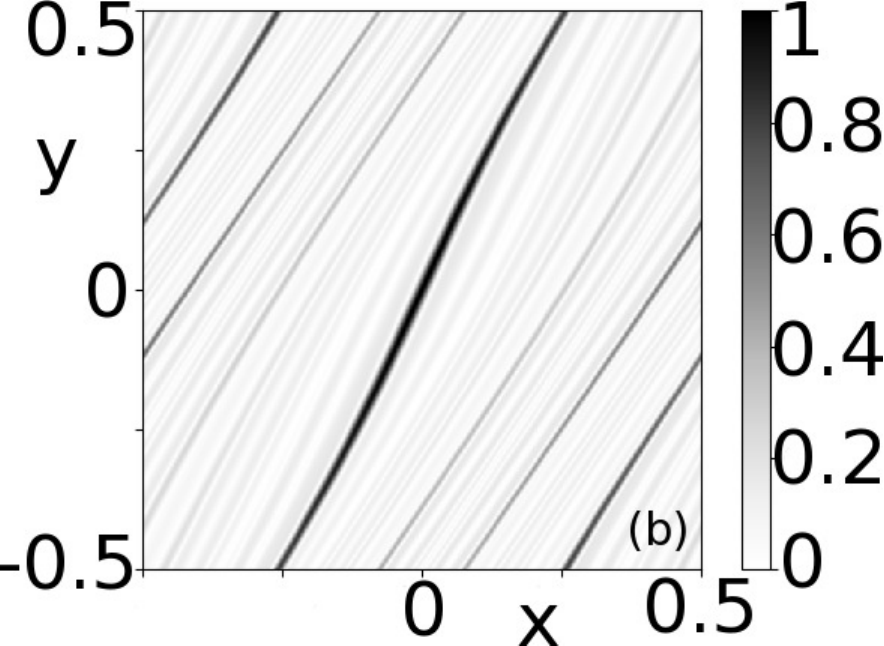}
}      
\caption{Magnitude of the second eigenfunction of the noisy Perron-Frobenius operator, 
  numerically evaluated for: (a) Bunimovich quarter stadium billiard (in the inset with the bouncing-ball orbit) with noise of amplitude $D=10^{-2}$, using 
  the Ulam matrix~(\ref{eq:ulam}) with $N=2^{14}$. Here $(\phi,\sin\theta)$ are the coordinates of the dynamics on the boundary of the billiard: 
  $\phi$ is the polar angle locating the bounce, and $\sin\theta$ is the angle of incidence with the normal to the boundary;  
(b) perturbed cat map (periodic point at the origin) with $\epsilon=0.1$, $\nu=1$, using the scheme~(\ref{FourOp}) with diffusivity $\Delta=5\times10^{-3}$ and $M=100$.}
\label{mainres}
\end{figure}

\begin{center} 
\begin{figure}[tbh!]
\includegraphics[width=0.26\textwidth]{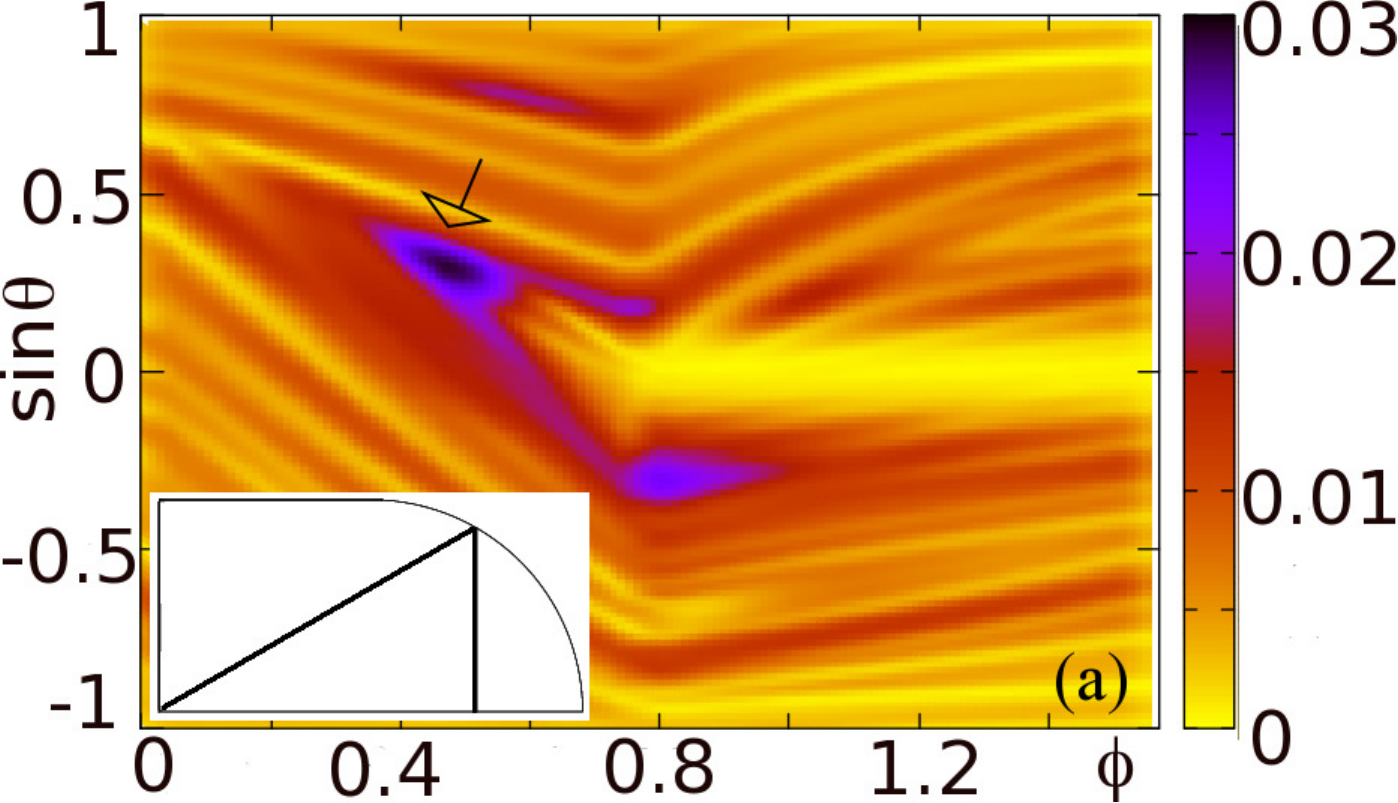}
\includegraphics[width=0.26\textwidth]{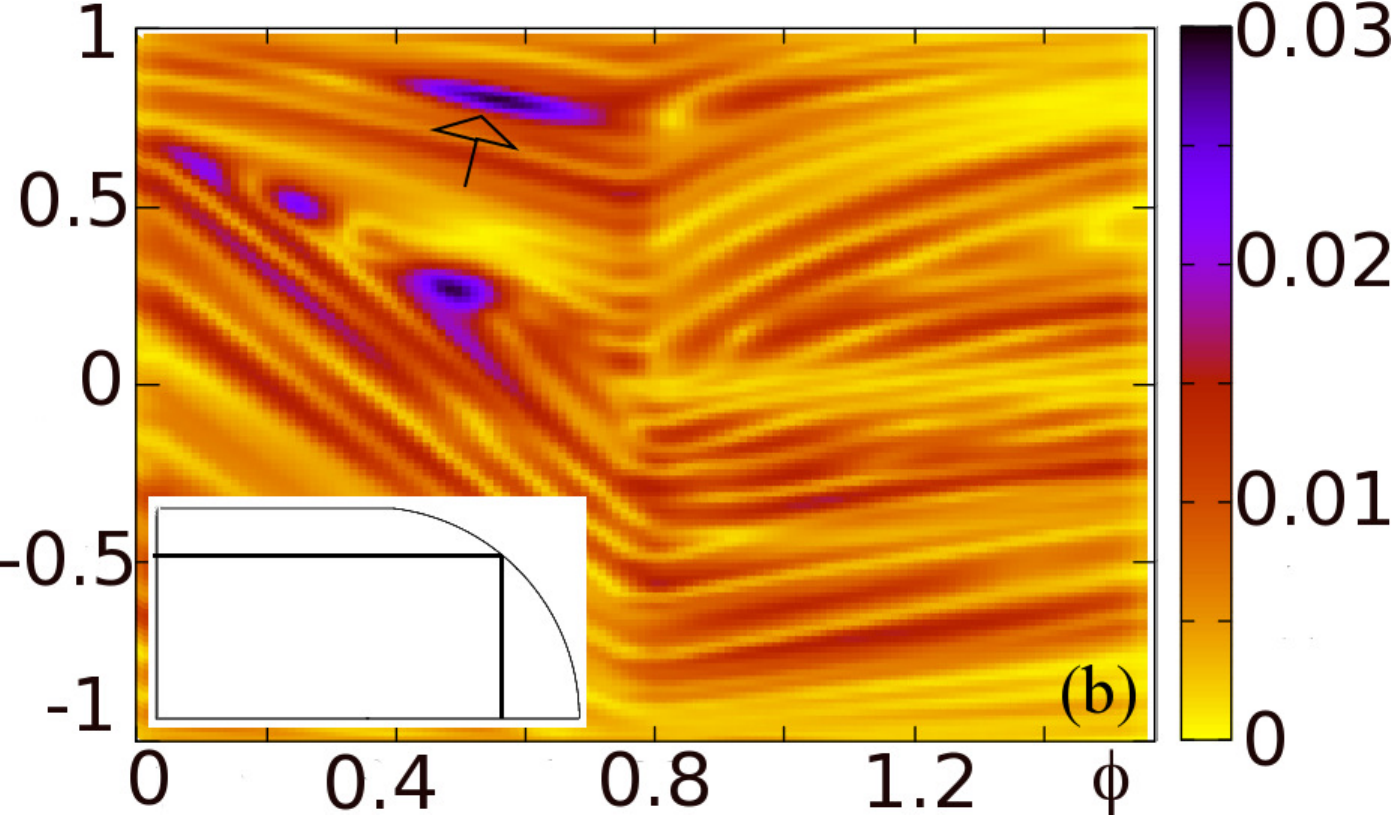}
\includegraphics[width=0.26\textwidth]{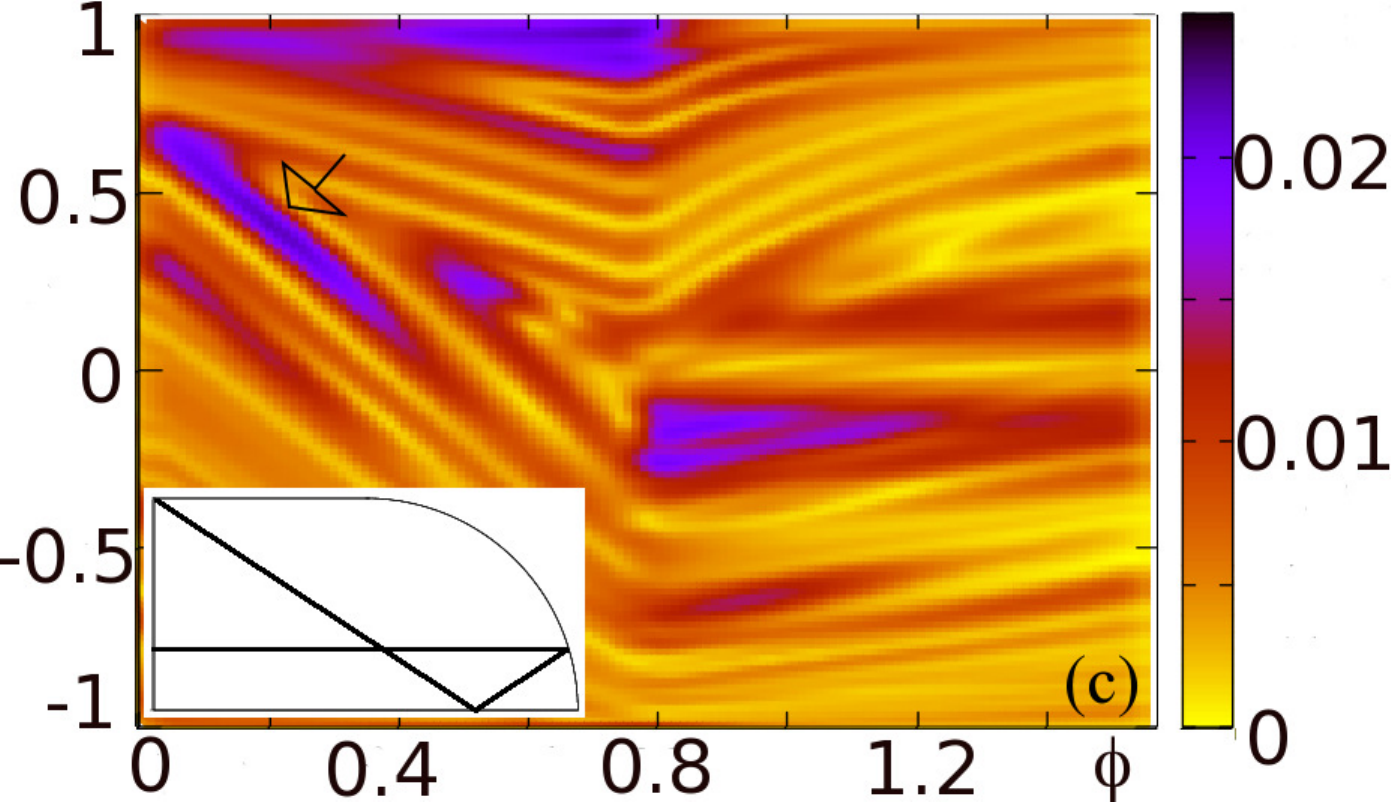}
\includegraphics[width=0.2\textwidth]{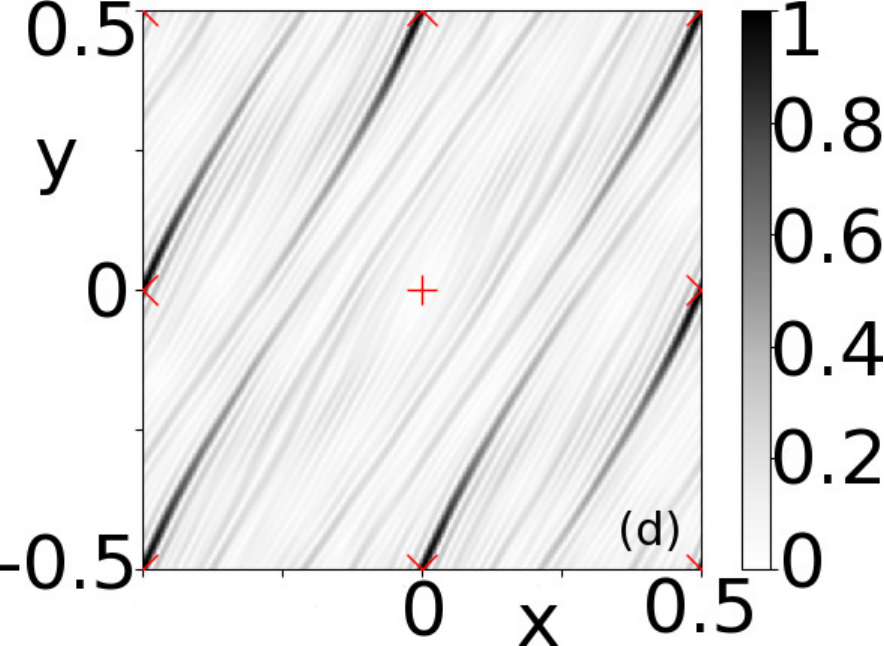}
\caption{Classical scars of short periodic orbits: 
%(pointed by arrows) in the eigenfunctions of the
%noisy Perron-Frobenius operator for the 
(a)-(c) Bunimovich quarter stadium billiard. Here
%with the corresponding periodic orbits in the insets.
the Ulam matrix~(\ref{eq:ulam}) has $N=2^{14}$ , and $D=10^{-2}$. (a) Bowtie orbit (corresponding scar pointed to by an arrow); 
%[for the noiseless billiard  $(\phi,\sin\theta)=(\frac{\pi}{6},\frac{1}{2})$];
(b) rectangular orbit; 
%[for the noiseless billiard  $(\phi,\sin\theta)\simeq(0.4,0.7)$];
(c) triangular orbit; 
%[for the noiseless billiard  $(\phi,\sin\theta)\simeq(0.15,0.3)$].} 
(d) scar of a period-3 orbit [marked by $(\times)$] of the perturbed cat map ($\epsilon=0.1, \nu=2$) on the unit torus, obtained as the second eigenfunction of the transfer matrix~(\ref{FourOp}), and $M=50$, $\Delta=10^{-5}$. The fixed point at the origin $(+)$ is `anti-scarred'. }
\label{ScarParade}
\end{figure}
\end{center}

\textit{Classical scars.} Figure~\ref{mainres} illustrates localization of
the subleading eigenfunctions of the noisy Perron-Frobenius operator near classical periodic orbits,
for the Bunimovich quarter stadium billiard~\cite{Bunim74}, as well as the cat map
perturbed with a nonlinear shear, 
$(x',y')= \left(x+y-\frac{\varepsilon}{\nu}\sin(2\nu\pi y),x+2y-\frac{\varepsilon}{\nu}\sin(2\nu\pi y)\right)$ mod 1 .
%(also called `Anosov map'). 
In analogy with the corresponding enhancement of probability density of quantum
eigenstates, we dub the observed phenomenon \textit{classical scarring}, 
with the caveat that, to present knowledge, while some mechanisms behind the formation of
scars are common in  classical and in quantum mechanics, others may differ between the two. 
%not all systems that
%exhibit quantum scarring must also feature classical scarring and vice versa.
The dynamics of the cat map is everywhere unstable (`hyperbolic')~\cite{AA68}, 
and the slowly-decaying eigenfunctions of the
Perron-Frobenius 
%(and Fokker-Planck) 
spectrum are striated along the 
unstable manifold~\cite{Haake00,BAgam00,Weber01,Manderfeld03}.
On the other hand, the stadium billiard is chaotic, ergodic~\cite{Bun79}, and has infinitely many unstable periodic orbits,
but it also possesses a family of marginally stable (`bouncing ball') orbits that give rise to 
 corresponding scars in the eigenfunctions of the quantized system~\cite{SuperBogo,HHBB10,Sieb93}. 
An example
of their classical counterpart is shown in Fig.~\ref{mainres}(a).
%Our results for the Bunimovich stadium billiard
As for localization arising from unstable orbits in the stadium, 
results in Fig.~\ref{ScarParade} suggest %show 
that the same scars recur in distinct subleading eigenfunctions, and, conversely,
that a single eigenfunction often displays several scars.
%In what follows,
%we provide details of our models and numerical schemes, and a mechanistic explanation for 
%the formation of classical scars in noisy chaotic dynamics.   
\onecolumngrid
\twocolumngrid      
\textit{Methodology.} The noisy Perron-Frobenius operator
%~(\ref{FP}) 
is projected onto a finite-dimensional vector space, and thus implemented as a finite matrix.
Previous literature warns us that the choice of the discretization
is crucial and may deeply affect the eigenspectrum beyond the leading 
eigenvalue in the linear map~\cite{Brini}. It has been established, on the other hand,
 that both nonlinear perturbations to linear maps on a torus, and 
background noise, increase the robustness of the numerically evaluated spectrum under certain conditions~\cite{BKL02}.      

The simplest discretization scheme is Ulam's method~\cite{Ulam}, that amounts to 
%building a uniform
%grid for 
subdividing 
the phase space into $N$ intervals ${\cal{M}}_i$ of equal area. 
%each being the support for a characteristic function.
The evolution operator is thus approximated with a $N\times N$ transfer matrix
whose entries $\Lmat_{ij}$ are the transition probabilities from  ${\cal{M}}_i$ to  ${\cal{M}}_j$
 %of entries
\begin{equation}
\Lmat_{ij} \,=\, \frac{\mu\left({\cal{M}}_i\bigcap f_\xi({\cal{M}}_j)\right)}
{\mu({\cal{M}}_i)}
%    \int_{{\cal{M}}_i} \! dx  \int_{{\cal{M}}_j} \!dy \,
%    \mathrm{e}^{-(y-f(x))^2/4D}
%\,.
\label{eq:ulam}
\end{equation}
%Instabilities have been reported that make Ulam's method in general unreliable for the
%computation of the whole spectrum of the evolution operator\cite{Brini}. 
%Each matrix element 
in one time step, where $f_\xi(\xpt)=f(\xpt)+\xipt$ is the noisy mapping,
while $\mu$ is the Lebesgue measure.
We use a known Monte Carlo method~\cite{ErmShep12} to estimate 
the non-symmetric transfer matrix $\Lmat_{ij}$, 
a (weighted) directed network~\cite{Netwks}, in today's parlance.
%: place $n_s$ points $x$ randomly in each region ${\cal{M}}_j$, map every $x$ once
%to $\xpt'=f(\xpt)+\xipt$, with $\xipt$ noise of amplitude $D$, thus count the $n_e$ iterates that land in ${\cal{M}}_i$, hence
%set  $[{\cal{L}}]_{ij}=n_e/n_s$.     
%That way, the non-unitary Perron-Frobenius %(or Fokker-Planck) 
%operator is Markov-approximated 
%by the non-symmetric transfer matrix $[{\cal{L}}]_{ij}$, %nowadays called
%a (weighted) directed network~\cite{Netwks}, in today's parlance.
A thorough study of stability and convergence of discretization algorithms  
has been reported elsewhere by the authors~\cite{Kens20}. 
%The evolution operator of the
%The Ulam matrix~(\ref{eq:ulam}) is used to approximate 
%and, in general, we cannot
%be positive about the accuracy of the single eigenvalues/eigenfunctions other than the natural measure.
 
\begin{figure}[tbh!]
\centerline{
%(a)
\includegraphics[width=3.8cm]{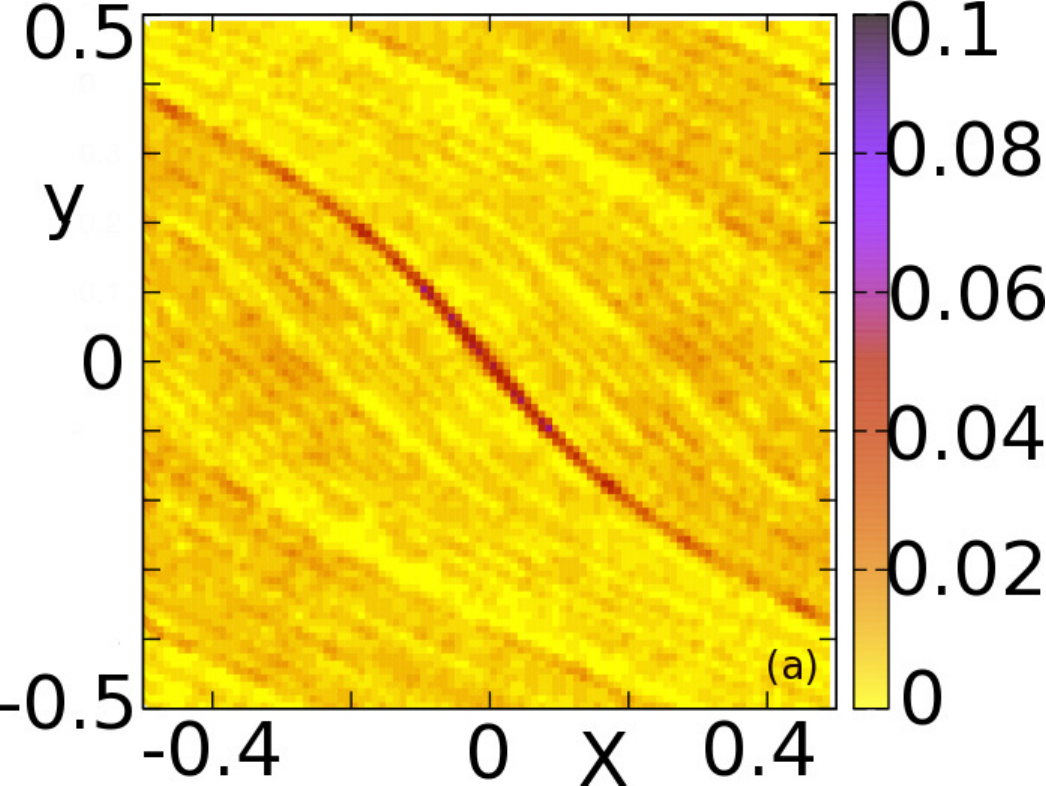}
%(b)
\includegraphics[width=3.8cm]{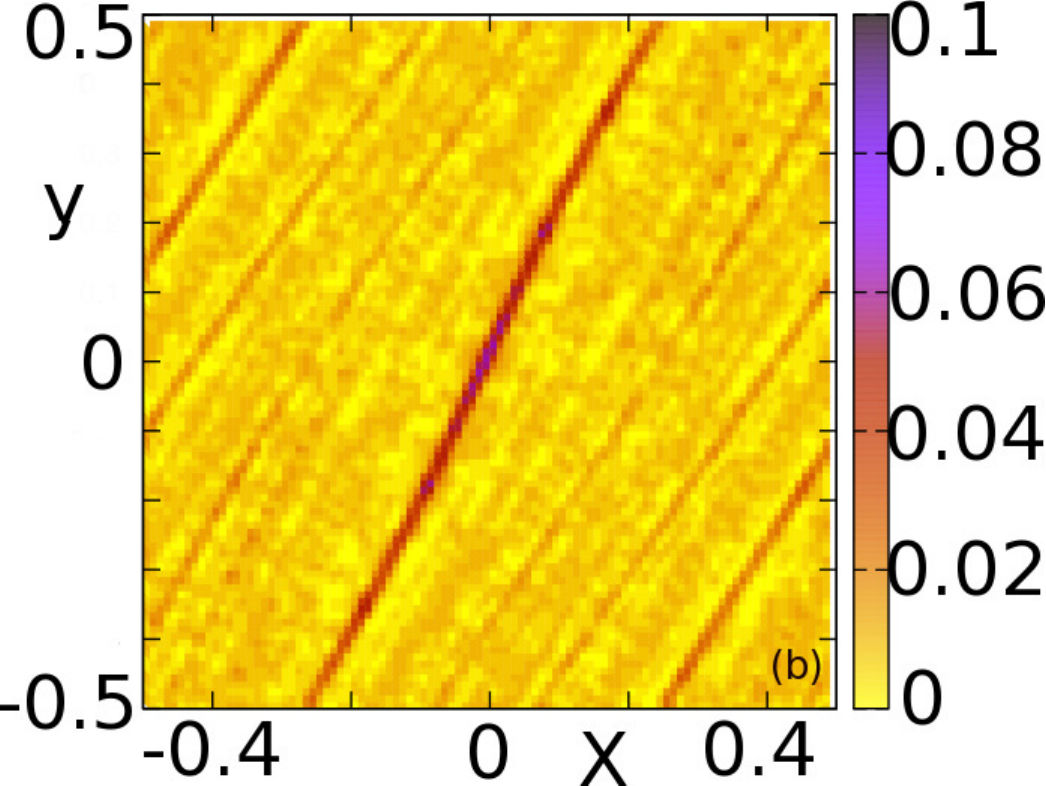}
} 
\vskip 0.1cm
\centerline{
%(c)
\includegraphics[width=3.8cm]{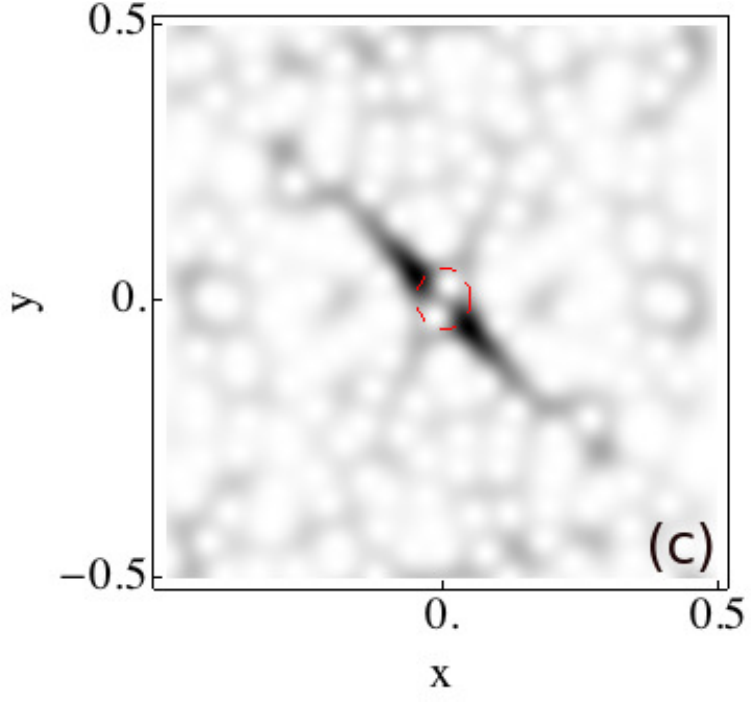}
%(d)
\hskip 0.7cm
\includegraphics[width=3.9cm]{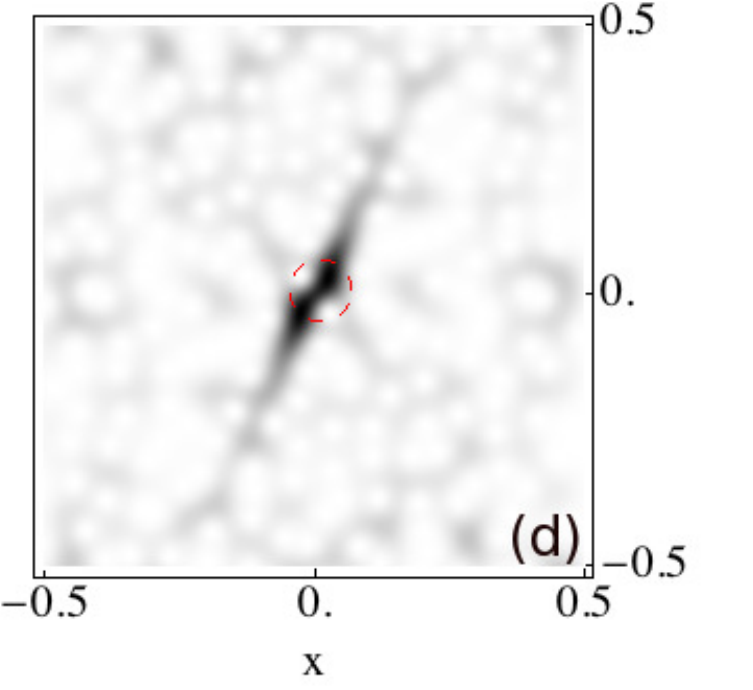}
}           
\caption{The perturbed cat map: magnitudes of the (a) left- and (b) right second eigenfunctions
of the Perron-Frobenius operator, realized through
the Ulam matrix~(\ref{eq:ulam}), with $N=10^4$, $D=10^{-2}$; 
Husimi distributions of (c) right- and (d) left eigenfunctions of the spectrum of the subunitary quantum
propagator coupled to a single-channel opening (dashed circle). Details of the
quantization in ref.~\cite{LRLK}.}
\label{analogy}
\end{figure}

We implement Ulam's scheme for the Bunimovich quarter stadium, where more sophisticated 
discretizations (e.g. Markov partitions) appear impractical. 
On the other hand, the perturbed cat map also
allows for an alternative realization of the transfer operator, by whose means we rule out
the possibility that the detected scarring be just a numerical artifact.           
%Here we mainly focus on
%the slowest-decaying eigenfunctions, and we want to make sure that the present
%observation of classical scarring is not just a numerical artifact.
%To that end, smooth basis functions are introduced for an alternative discretization
%of ${\cal{L}}$ for the Anosov maps. 
Since the dynamics of the cat map lives on the unit torus, a basis of smooth, periodic functions is
suitable for the evolution operator, that can be defined in Fourier space as~\cite{ThiffChild}
\begin{equation}
{\cal{L}}_\Delta \rho(\xpt) = \int \sum_{k_x,k_y}^M \mathrm{e}^{2\pi i \kvec\cdot (f^{-1}(\xpt)-\xpt_0)- \Delta\kvec^2}\rho(\xpt_0)d^2x_0
\,.
\label{FourOp}
\end{equation}
Here the diffusivity $\Delta$ is equivalent to the variance $D$ of the
random variable $\xipt$ defined above in the Langevin picture. 
The spectrum of the operator in this basis is robust under perturbations~\cite{Kens20}
(e.g. dimension of the transfer matrix, noise amplitude, nonlinearity of the perturbed cat map), and
classical scarring is consistently detected in the second eigenfunction of the
spectrum: the one in Fig.~\ref{mainres}(b) is computed with the Fourier basis of Eq.~(\ref{FourOp}),
while, for the same map, the second eigenfunction of the Ulam matrix~(\ref{eq:ulam}) 
%${\cal{L}}_{ij}$
is displayed in Fig.~\ref{analogy}(b).

%The Fourier discretization~(\ref{FourOp}) is not suited for the Bunimovich stadium billiard,
%instead, for its phase space is only periodic in the azimuthal angle, but not in the momentum.

\textit{Analogy with quantum scars.} The route to understand classical scarring begins by 
comparing it directly with its quantum counterpart.
However, the propagator $U^t$, that regulates the evolution of quantum cat maps,  
is unitary~\cite{Creagh95}, unlike 
%, when discretized using smooth basis functions
 %That calls for a premise regarding 
%the evolution operators involved: the non-unitarity of the Perron-Frobenius ${\cal{L}}$ 
%or the Fokker-Planck operator ${\cal{L}}_{FP}$ 
%However, 
our realizations of the noisy Perron-Frobenius operator ${\cal{L}}^t$, 
%are non-unitary, 
%of  Eq.~(\ref{PF})
%(\ref{FP}) respectively 
%in classical dynamics, versus the unitarity of 
%whereas 
%unlike the quantum propagator $U^t$, that possesses group properties
%and a  
%is unitary, and that 
and that constitutes a clear asymmetry in our quest for classical-to-quantum correspondence.               
Breaking the unitarity of $U^t$ by coupling the quantized cat map to an opening 
is the simplest way to restore the symmetry between classical and quantum evolution.
The result is exemplified in Fig.~\ref{analogy}, that features scars around the 
periodic orbit at the origin of the phase space, in both noisy classical and quantum  
cat maps. In  the classical setting, the areas of enhanced probability density are striated along the stable- 
(left eigenfunction of ${\cal{L}}$) 
or unstable (right eigenfunction of ${\cal{L}}$) manifold, that emanates from
the periodic orbit located at the origin. The correspondence left/right eigenfunction-stable/unstable
manifold is less straightforward for the open quantum map~\cite{LRLK},
but still one-to-one.

\begin{figure}[tbh!]
\centerline{
%(a)
%\begin{center}
\includegraphics[width=4.8cm]{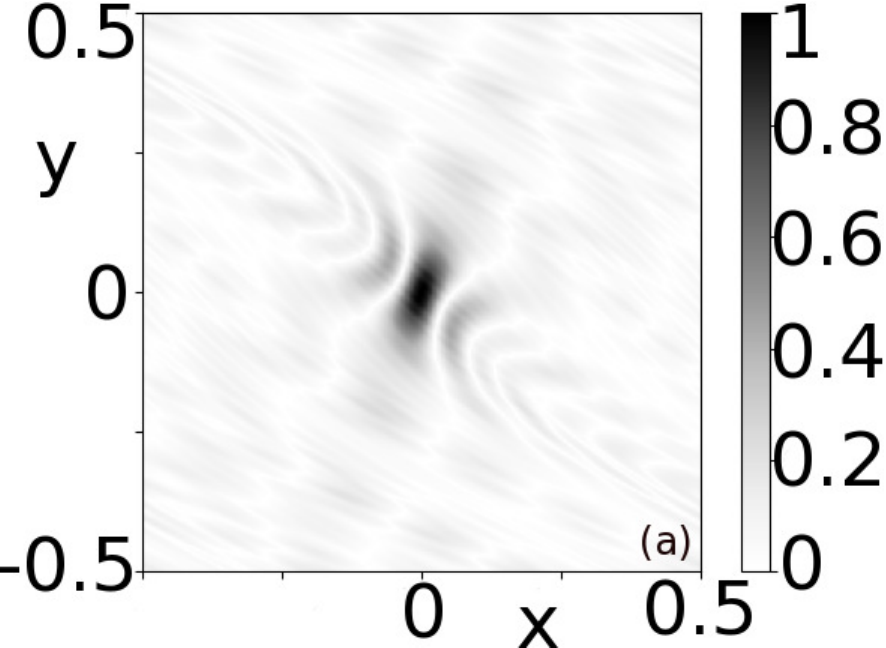}
%(b)
%\vskip 0.1cm
\includegraphics[width=3.6cm]{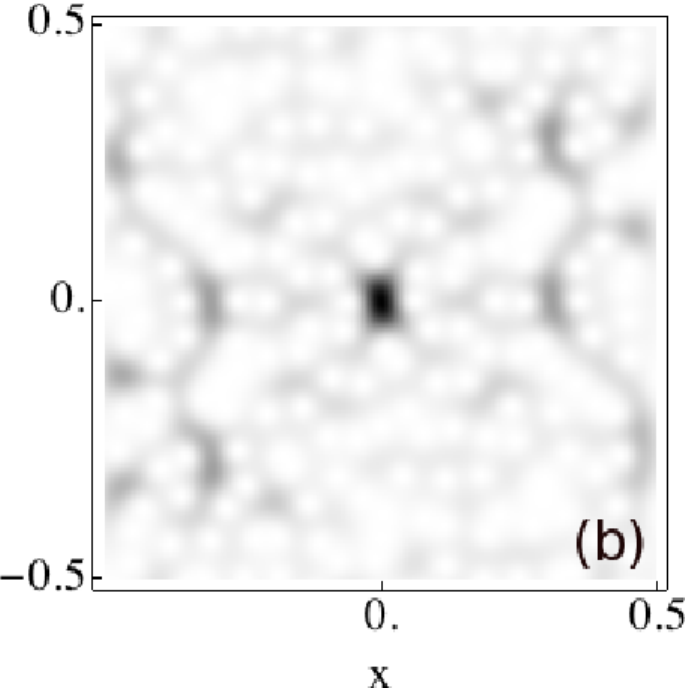}
}      
%\end{center}
\caption{(a) Unwrapping: the operator~(\ref{FourKoop}) is applied for $t=3$ iterations to
 the second eigenfunction of the noisy Perron-Frobenius operator for the perturbed cat map
 [parameters as in Fig.~\ref{LyapDemo}(a)];  
(b) Husimi distribution of a scarred eigenfunction of the unitary quantum propagator of the 
same map.}  
\label{unwrap}
\end{figure}

Conversely, we may assimilate the classical scars to
the original quantum scars of closed chaotic systems, 
where the propagation is unitary.
Quantum scars of a unitary propagator are typically concentrated 
around a periodic orbit with no elongations on the manifolds, as a result of the 
unitary evolution.
Using the known technique of eigenfunctions unwrapping~\cite{Froyl},
we map a right eigenfunction of the noisy ${\cal{L}}$  backward in time by means of the  
adjoint (`Koopman') evolution operator, whose noiseless definition %in the absence of noise 
reads 
%[note the change in the variable of integration with respect to Eq~(\ref{FP})]
\begin{equation}
\left[{\cal{L}}^t\right]^\dagger\rho(\xpt) = \int dx_0\,\delta\left(\xpt_0-f^t(\xpt)\right)\rho(\xpt_0) = 
\rho\left(f^t(\xpt)\right)
\,.
\label{FPdag}
\end{equation}
When noise is included and the Fourier representation~(\ref{FourOp}) is used for
${\cal{L}}_\Delta$, 
the adjoint evolution operator is 
%can be 
%the transpose of the Ulam transfer matrix~(\ref{eq:ulam}), or,
%for 
\begin{equation}
{\cal{L}}_\Delta^\dagger \rho(\xpt) = 
 \sum_{k_x,k_y}^M \mathrm{e}^{2\pi i \kvec\cdot f(\xpt)-\Delta \kvec^2}
\hat{\rho}_\kvec
 \,,
 \label{FourKoop}
 \end{equation}
 where the $\hat{\rho}_\kvec$'s are the Fourier coefficients of the density $\rho(\xpt)$.  
Repeated adjoint mapping rids the eigenfunction of the striation along the
unstable manifold, and regularizes it, %The support of the
%unwrapped density is partitioned into sets   
until the outcome of Eq.~(\ref{FourKoop}) is almost invariant under both
forward and backward iterations. The so-obtained unwrapped eigenfunction [Fig.~\ref{unwrap}(a)]
%As shown in Fig.~\ref{unwrap}, one of these sets is the scar, 
now closely resembles 
its quantum analog  [Fig.~\ref{unwrap}(b)].
%the unwrapped eigenfunction of ${\cal{L}}_{FP}$ closely resembles
%a quantum scarred eigenstate.

\textit{Origin of classical scars.} We now examine the local action of the classical 
Perron-Frobenius operator~(\ref{PF}) on phase-space densities, in order to 
gain some insight on the
mechanism behind classical scarring in uniformly hyperbolic systems such as the cat map:
\begin{equation}
{\cal{L}}^t\rho = \int dx_0\,\delta\left(\xpt-f^t(\xpt_0)\right)\rho(\xpt_0) =
\sum_{\ypt=f^{-t}(\xpt)} \frac{\rho(\ypt)}{|\det J^t(\ypt)|}
\,,
\label{PFdvp}
\end{equation}
where $J_{ij}^t(\ypt)=\frac{\partial_i f^t(\ypt)}{\partial y_j}$ is the Jacobian of the flow. 
Now restrict the analysis to the unstable manifold, where densities are
stretched and squished. If the map is 2D, the unstable manifold is a 1D curve characterized by an arc length $s(\ypt)$, and
we indicate the restricted dynamics with $f_u(s)$, the corresponding evolution operator with ${\cal{L}}^t_u$, and a density on the manifold with $\rho_u(s)$. 
The mapping takes the form  
\begin{equation}
{\cal{L}}^t_u\rho_u(s) \propto \prod_{k=0}^{t-1} \frac{1}{\left|\left[f^{k}_u(s)\right]'\right|} \rho_u(s)
= \mathrm{e}^{-\Lambda(\ypt,t)t}\rho_u(s)
\,,
\label{unst_ev}
\end{equation}
where $\Lambda(\ypt,t)$ is the finite-time Lyapunov exponent~\cite{fuji83} 
of the map $f^t(\xpt)$, that is 
the rate of exponential divergence of nearby trajectories within the time~$t$.
The last equality in Eq.~(\ref{unst_ev}) stems from the fact that the expanding rate of
$f_u(s)$ is the stability multiplier of $f(\mathbf{x})$. 
%orbit the manifold emanates from.  
On the other hand, assume  that the relaxation of $\rho(\ypt)$ towards equilibrium is well described by
a truncation of the expansion~(\ref{Lexp}),  
%for the evolution of a density
%onto the unstable manifold,
whose slowest-decaying term is almost entirely supported on 
the unstable manifold: 
%\footnote{this is a projection onto the unstable manifold, but then
%ont the right-hand side we have the first two terms of the expansion. What was the effect of the projection?}:
\begin{equation}
{\cal{L}}^t\rho(\ypt) \simeq a_0 + a_1\mathrm{e}^{-\gamma_1t}\phi_1(\ypt) \sim  
a_0 + \mathrm{e}^{-\Lambda(\ypt,t)t}\rho_u(s) 
\,.
\label{exp2terms}
\end{equation}
%Let the projected density $\rho_u(y)=\rho_0$ be uniform on the unstable manifold.
%Now we can say that 
Since  the evolution~(\ref{unst_ev}) of a density on the unstable manifold does not depend on the initial condition $\rho_u(s)$, 
we infer that
$\Lambda(\ypt_1,t)>\Lambda(\ypt_2,t)\Rightarrow |\phi_1(\ypt_1)|<|\phi_1(\ypt_2)|$,
and thus in general the probability density of the second eigenfunction of the spectrum   
along the unstable manifold is ruled by the finite-time Lyapunov exponent:  
the lesser instability, the higher the (magnitude of the) density~\cite{HaynVann05,ThiffLect08,Haller15}. 
%This analysis is particularly relevant, given that the second eigenfunction of the
%Perron-Frobenius (and Fokker-Planck) spectrum is striated along the 
%unstable manifold~\cite{BAgam00} [Fig.~\ref{LyapDemo}(b)]. 
%In making this argument
%we have assumed no degeneracy in the second eigenvalue, which is a fact for the
%Anosov maps of our choice.
 As it can be inferred from Eq.~(\ref{exp2terms}),
the Lyapunov exponent is to be evaluated over a time $t\simeq\gamma_1^{-1}$,
thus of the order of the decay time of the eigenfunction $\phi_1(\mathbf{x})$. %\footnote{Is that true?}. 
%Where does noise fit in this picture? 
Densities stretch out along the unstable manifold,
while they are contracted along the stable manifold. Asymptotically, the
compression makes them infinitesimally thin, but %physical 
noise counters
that effect, and fattens densities along the unstable manifold. As a result, scarring also
becomes apparent by visual inspection.     

\begin{figure}[tbh!]
\centerline{
%(a)
\includegraphics[width=4.2cm]{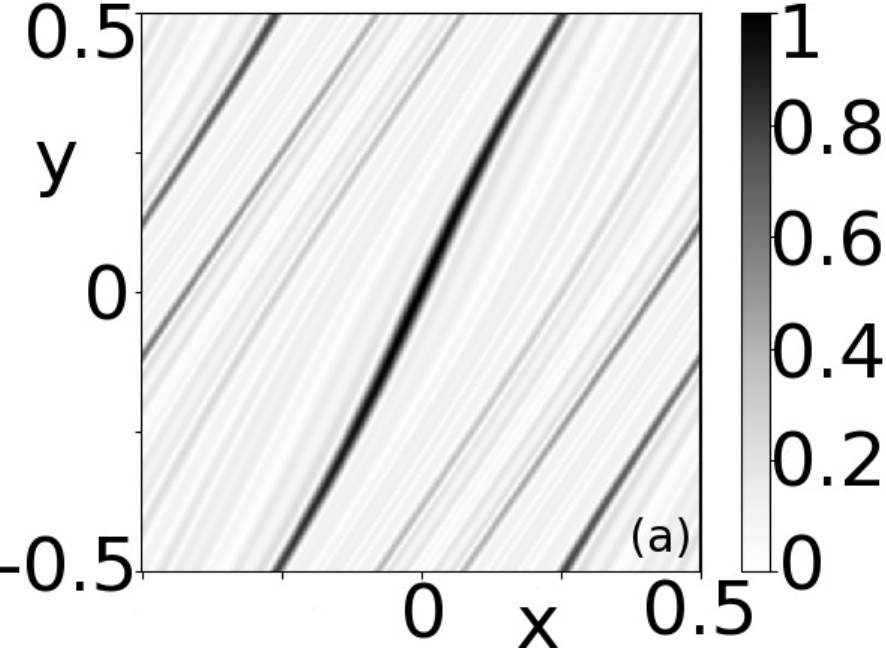}
%(b)
\includegraphics[width=4.2cm]{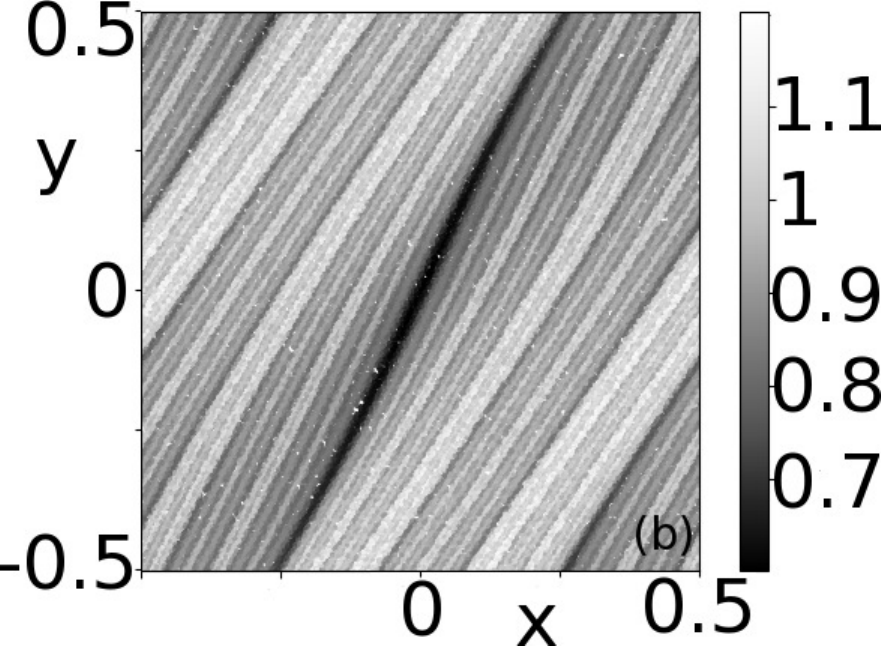}
}      
%\centerline{
%(c)
%\includegraphics[width=4.4cm]{Ken3FoldEig.png}
%(d)
%\includegraphics[width=4.4cm]{Ken3FoldLyapDist.png}
%}      
\caption{%(a) %Detail of the Perron-Frobenius spectrum of a perturbed cat map, pointing
%out three degenerate eigenvalues corresponding to as many distinct scars;
%(b) unstable manifold of the same map; 
(a) The second eigenfunction of the noisy Perron-Frobenius operator~(\ref{FourOp}), 
 numerically evaluated for a perturbed cat map with $\Delta=10^{-2}$ and cutoff $M=100$;
(b) the phase-space distribution of the finite-time Lyapunov exponents ($t=5$, sampling: $10^5$ initial points) for the same map.}  
\label{LyapDemo}
\end{figure}
Figure~\ref{LyapDemo} supports this hypothesis: the second eigenfunction of the  
noisy Perron-Frobenius operator is shown for the perturbed cat map, and it is localized
along the unstable manifold that emanates from the origin. 
%with a modified perturbation, so as to 
%exhibit a scar on a period-three orbit, besides the scar at the origin. 
On the other hand, the numerically computed phase-space 
distribution of the finite-time Lyapunov exponent displays a suppression pattern that nearly 
overlaps with the scar.%\footnote{and of the unstable manifold~\cite{Haller15}}.

%We provide further evidence for the above argument by 
\textit{Power spectra.} Scarring can be quantified by
studying the power spectrum~\cite{OttSpect96}
\begin{equation}
S(\omega) = \frac{1}{\sqrt{T}}\sum_{t=0}^T {\cal{C}}(t)\mathrm{e}^{2\pi i \omega t/T}
\label{Somega}
\end{equation}
of a Gaussian density (the classical analog of a wave packet), that gradually decays into a 
uniform phase-space distribution, as the 
%Fokker-Planck 
evolution operator is applied. Here $ {\cal{C}}(t)$ is the autocorrelation function of the 
density [defined in Eq.~(\ref{Capprox})], while $T$ is the length of the time series. 
 %In the discretized 
%picture, the autocorrelation functi 
\begin{figure}[tbh!]
\centerline{
\includegraphics[width=3.6cm]{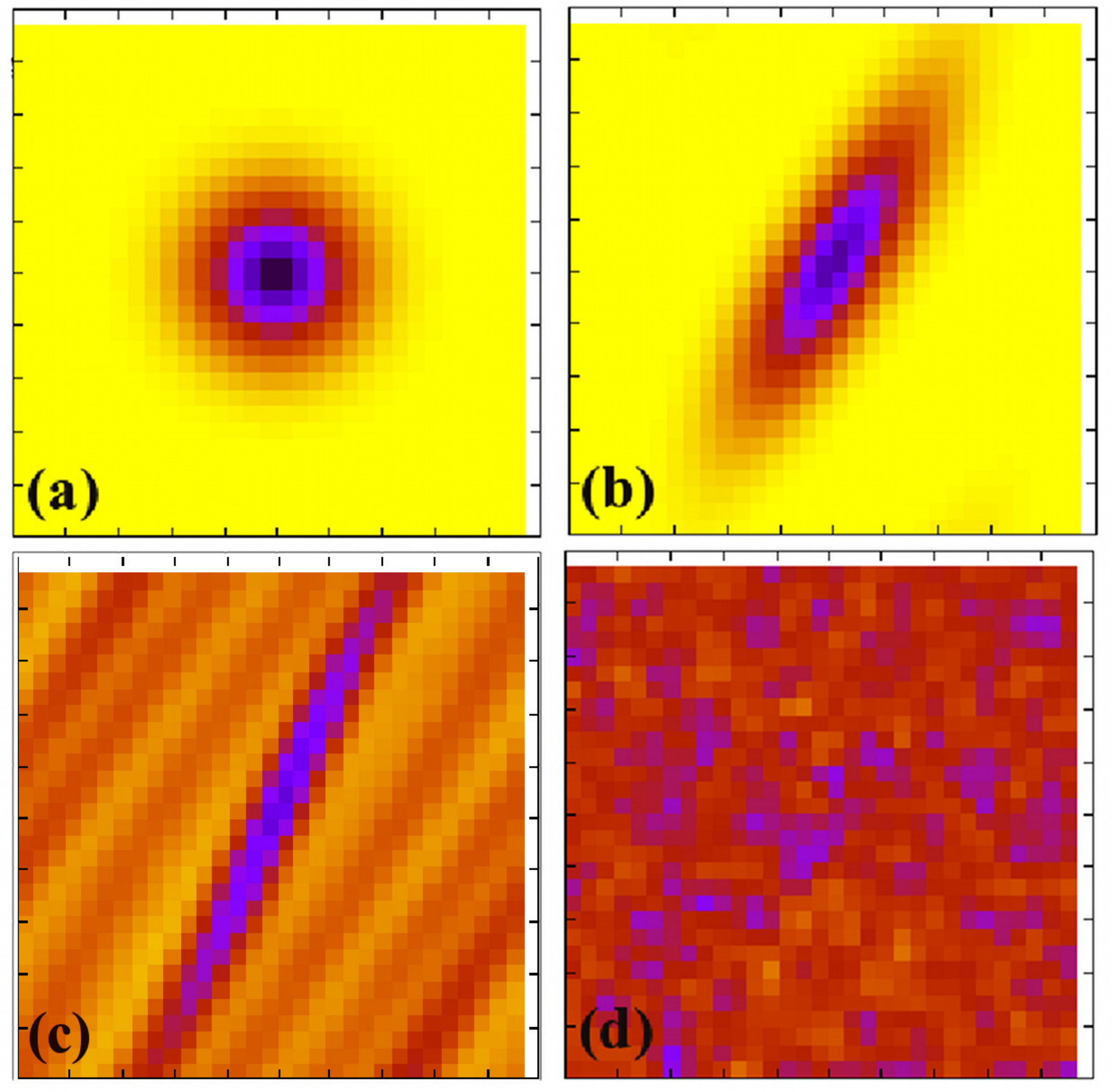}
%(a)
\includegraphics[width=5.cm]{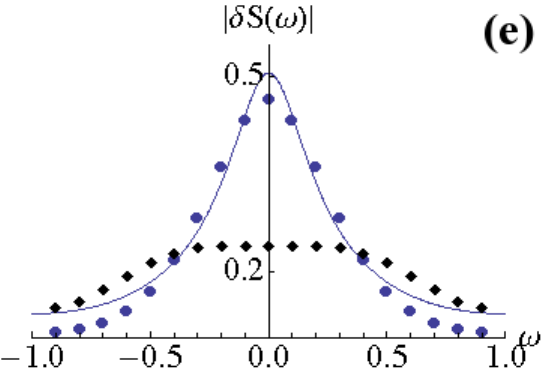}
%(b)
%
}      
%\centerline{
%(c)
%\includegraphics[width=4.4cm]{Ken3FoldEig.png}
%(d)
%\includegraphics[width=4.4cm]{Ken3FoldLyapDist.png}
%}      
\caption{(a) An initial Gaussian density centered at the fixed point of the perturbed cat map is
mapped by the Ulam matrix~(\ref{eq:ulam}); %$[{\cal{L}}]_{ij}$; 
(b) t=1 iterations; (c) t=5; (d) t=20.
(e) Magnitude of the power spectrum of the autocorrelation function minus the steady state: 
numerics for (dots) the density in (a)-(d), (diamonds) the same initial density centered at random in the
phase space, (solid line) prediction~(\ref{predspec}).}  
\label{PSpec}
\end{figure}
Figure~\ref{PSpec} shows the outcome of the numerical experiment. 
A fast decay of ${\cal{C}}(t)$ occurs if we place the initial density 
at random in the phase space, resulting in a flat power spectrum. 
Instead, centering the initial distribution
around the scarred fixed point [Fig.~\ref{PSpec}(a)] produces a slower decay in the 
autocorrelation function, and a peaked power spectrum.
The quantum analog of $|S(\omega)|$ is the local density of
states, whose energy-dependent, peaked envelope is 
a well-known signature of scarring~\cite{KapHel98}.   
The power spectrum is related to the second eigenfunction of the transfer operator in the
following way. Truncating the expansion~(\ref{Lexp}) at the first order, 
the autocorrelation function is estimated as 
\begin{equation}
{\cal{C}}(t) =  \frac{{\cal{\rho}}_0^T\left[{\cal{L}}^t{\cal{\rho}}_0\right]}{{\cal{\rho}}_0^T{\cal{\rho}}_0}
\approx  c_0 + c_1\mathrm{e}^{-\gamma_1 t}
\,,
\label{Capprox}
\end{equation}  
 for some $c_0, c_1$.
 The discrete Fourier transform~(\ref{Somega}) of Eq.~(\ref{Capprox}) then yields a delta function
 from the asymptotic overlap $c_0$ of the evolved density with the natural measure, plus the 
 actual power spectrum of the exponential, 
 \begin{equation}
\delta S(\omega) \propto  \frac{1}{1-\mathrm{e}^{i\omega T/2\pi-\gamma_1t}}%\mathrm{e}^{-\gamma_1t}
\,.
\label{predspec}  
\end{equation}
This approximation [solid line in Fig.~(\ref{PSpec})], with $\gamma_1$ determined from the diagonalization
of the transfer matrix~(\ref{eq:ulam}), %$[{\cal{L}}]_{ij}$,
 shows close agreement with
% the power spectrum
%of the autocorrelation function computed numerically
the direct numerical computation. Using the results of  Eq.~(\ref{exp2terms})
and Fig.~\ref{LyapDemo}, one could replace $\gamma_1$ in Eq.~(\ref{predspec}) with the minimum
finite-time Lyapunov exponent evaluated at $t\simeq\gamma_1^{-1}$.
%\footnote{what about the finite-time Lyapunov exponent? May I
%replace $\gamma_1$ with $\Lambda_u^{(min)}$? But where is the noise?}.  

\textit{Conclusion.} We have reported the observation of classical scars, 
that is enhancement of probability density  
near periodic orbits, in the eigenfunctions of a noisy
evolution operator of chaotic systems.   
We have detected scars in two model systems: perturbed cat maps, and 
the Bunimovich stadium billiard, both with background noise.
For the cat map, the observed localization is ascribed to the inhomogeneity of 
instabilities near the periodic orbit of interest, on a time scale
consistent with the decay rate of the second eigenfunction of the
evolution operator, that is also the rate of correlation decay.
In support of this argument,
we have compared the second eigenfunction with the distribution of 
finite-time Lyapunov exponents of the dynamical system in exam.
The inevitable presence of noise does not alter this mechanism, but 
merely thickens the scars by its smearing action along the unstable  
manifolds [cf. Figs.~\ref{mainres}(b) and~\ref{LyapDemo}(a)]. 
Although we do not claim a one-to-one correspondence
between classical and quantum scarring, we have pointed out
apparent similarities in their phase-space patterns (provided the same 
symmetries) and power spectra, with the important difference of
the time scales associated to the relevant instabilities. 
%quantum 
%scars have been characterized by local instabilities, while here
%we identify the Lyapunov exponents evaluated on a short but finite time,
%as the ruling classical observables.      
It is then natural to suppose that the mechanism at the origin of 
classical scars must also play a role in the formation of their 
quantum counterparts, in the same spirit as %recent work
%on 
classical dynamical localization~\cite{ClasDynLoc},
or  phase-space localization in open systems~\cite{Ketz18,Ketz19}.    
%Results in that direction have appeared 
%in the last decade,
%for example,
%in reinterpreting dynamical localization~\cite{ClasDynLoc} classically,
%or, most lately, to account for phase-space localization in open systems~\cite{Ketz18,Ketz19}.   
  
%The scarring is 
%ascribed to the striation of the slowest-decaying 
%eigenfunctions of the Fokker-Planck operator along the 
%unstable manifold, as well as to their non-uniform distribution of 
%finite-time Lyapunov exponents. %a chart of instabilities in the phase-space.
%On the other hand, noise plays a non-negligible role, by balancing the contraction along the
%stable directions, and giving the scars a finite width.
%Although we do not claim a one-to-one correspondence
%between classical and quantum scarring,
%it is natural at this point to suppose that the presence of a
%gradient in the distribution of instabilities near a periodic orbit
%of a chaotic system also plays a role in
%the formation of quantum scars.

\textit{Acknowledgments-} The authors are supported by 
NSF China (Grant No. 11750110416-1601190090) and 
JSPS KAKENHI (Grants No. 15H03701 and No. 17K05583). DL thanks
S. Pascazio and R. Bellotti for hospitality at INFN Bari, and
ReCaS Bari for computational resources.

\end{document}